\documentclass[aps,prb,amsfonts,amssymb,twocolumn,amsmath,preprintnumbers,nofootinbib,floatfix,showpacs]{revtex4}
\usepackage{amsmath,bm,amsfonts,amssymb}
\usepackage[dvips]{graphics}
\usepackage{graphicx}

\begin{document}

\title{Influence of spin filtering and spin mixing on
the subgap structure of I-V characteristics in superconducting
quantum point contact.}
\author{I. V. Bobkova}
\email[Electronic address: ]{bobkova@issp.ac.ru}
\affiliation{Institute of Solid State Physics, Chernogolovka,
Moscow reg., 142432 Russia}
\author{A. M. Bobkov}
\affiliation{Institute of Solid State Physics, Chernogolovka,
Moscow reg., 142432 Russia}

\date{\today}

\begin{abstract}
The effect of spin filtering and spin mixing on the dc electric
current for voltage biased magnetic quantum point contact with
superconducting leads is theoretically studied. The I-V
characteristics are calculated for the whole range of spin
filtering and spin mixing parameters. It is found that with
increasing of spin filtering the subharmonic step structure of the
dc electric current, typical for low-transparency junction and
junction without considerable spin filtering qualitatively
changes. In the lower voltage region and for small enough spin
mixing the peak structure arises. When spin mixing increases the
peak subgap structure evolves to the step structure. The voltages
where subharmonic gap features are located are found to be
sensitive to the value of spin filtering. The positions of peaks
and steps are calculated analytically and the evolution of the
subgap structure from well-known tunnel limit to the large spin
filtering case is explained in terms of multiple Andreev
reflection (MAR) processes. In particular, it is found that for
large spin filtering the subgap feature at $eV_k$ arises from
$2k^{\rm th}$ and $(2k\pm 1)^{\rm th}$ order MAR processes, while
in the tunnel limit the step at $eV_n$ is known to result from
$n^{\rm th}$ order MAR process.
\end{abstract}
\pacs{74.45.+c, 74.50.+r}

\maketitle

\section{Introduction}

In contrast to the voltage-biased junctions between two normal
metallic leads, the I-V characteristics of superconducting weak
links can be highly nonlinear in the subgap region $eV \lesssim
2\Delta$, where $\Delta$ is the superconducting order parameter in
the leads. This subharmonic gap structure is explained by multiple
Andreev reflection (MAR) and has been investigated in details in
various mesoscopic systems including superconductor/normal
metal/superconductor (SNS) plane junctions,
superconductor/insulator/superconductor (SIS) junctions and
quantum point contacts both theoretically
\cite{Klapwijk82,Octavio8388,Arnold,Shumeiko95,Averin95,
Cuevas96,Shumeiko01,Bardas97,
Zaitsev98,Shumeiko02,Bezuglyi99,Bezuglyi00,Kupriyanov03,Cuevas05}
and experimentally
\cite{vanderPost94,Scheer97,Scheer98,Ludoph00,Kutchinsky97,Hoss00}.
The differential conductance $dJ/dV$ of SNS, SIS and nonmagnetic
quantum point contacts exhibits a number of peaks at
$eV=2\Delta/n$ in the limiting regimes of short ballistic and
diffusive and long (in comparison with the superconducting
coherence length $\xi$) diffusive weak links. In the intermediate
regime $\xi \sim d$, when the interplay between proximity effect
and MARs takes place, the conductance subgap structure of SNS
voltage-biased diffusive junction modifies and has an additional
maximum at roughly $eV \sim \Delta + \Delta_g$. Here $\Delta_g$ is
a minigap in the equilibrium density of states in the normal
region arising due to the proximity effect \cite{Cuevas05}. In
addition, it was theoretically predicted that in long ballistic
SNS junctions coherence effects give rise to resonant structures
in the dc current due to Andreev quantization \cite{Shumeiko01}.

On the other hand, ferromagnetic weak links possess additional
characteristic parameters: spin filtering and spin mixing. Spin
filtering is the difference between transmission probabilities for
spin up and down electrons $D_\uparrow$ and $D_\downarrow$, which,
in particular, offers an opportunity to generate and control spin
currents by driving conducting electrons. Of course, this "spin
filtering" effect takes place in a spin-active junction between
two normal leads, but the corresponding spin current is linear in
the bias voltage. Superconducting hybrid structures with
ferromagnetic weak links give the possibility to obtain highly
nonlinear I-V characteristics, what can be essential for
manipulating spin currents in spintronic devices. By now there are
only several theoretical papers devoted to the investigation of
voltage-biased superconductor/ferromagnet/superconductor (SFS)
quantum point contacts. A. Martin-Rodero {\it
et.al.}\cite{Cuevas01} were the first to calculate the electric
current in SFS quantum point contact in the presence of spin
filtering effect $D_\uparrow \neq D_\downarrow$ and found that the
positions of subharmonic gap features in the I-V characteristics
are shifted as compared with the nonmagnetic case and located at
$eV_n \approx \Delta/(\sqrt 2 n)$ for $D_\uparrow = 1$ and
$D_\downarrow \ll 1$. Further the influence of another generic
property of spin-active interface called "spin mixing" on the
voltage biased electric current has been theoretically studied.
The essence of "spin mixing" is that spin up and spin down
electrons acquire different phases upon transmission or reflection
by a spin-active interface. The relative phase shift, so called
"spin mixing angle" $\Theta$ does not manifests itself in the I-V
characteristics of the junctions with normal leads, but results in
the shift of subgap step positions to $eV_n=\Delta(1 + \cos
\Theta)/n$ in the zero-temperature dc electric current of SFS
junction even without spin filtering \cite{Fogelstrom02}. Further
the simultaneous influence of spin filtering and spin mixing on
the dc electric and spin currents has been analytically studied
for the case of low-transparency SFS junction
$D_\uparrow,~D_\downarrow \ll 1$ \cite{Bobkova06}. It was found
that the spin mixing results in the splitting of subgap electric
current step positions at non-zero temperature leading to
$eV_n=\Delta(1 \pm \cos \Theta)/n$. The dc spin current also
manifests subgap structure of the same type in the
low-transparency limit, but it shows the odd-even effect: only the
steps corresponding to odd values of $n$ survive in the I-V
characteristics of spin current. It is worth to note  that
although this analytical analysis gives correct positions of the
spin current steps, the step height is considerably overestimated
compared to the non-perturbative numerical calculation except for
very tunnel limit $D_{\uparrow,\downarrow} \lll 1$
\cite{bobkovyunp}.

The most strongly non-linear and rich I-V characteristics can be
obtained in higher transmission SFS junctions. Such strongly
non-linear I-V characteristics for spin current have been
numerically calculated very recently\cite{Sauls06}. The present
paper is devoted to investigation of the dc electric current in
the voltage biased SFS quantum point contact for arbitrary spin
filtering and spin mixing parameters. For quantum point contact
sharp subgap features in the electric current only exist when at
least one of transparencies $D_{\uparrow,\downarrow}$ is low
enough. So it makes sense to focus on the two cases:
$D_{\uparrow,\downarrow} \ll 1$ and $D_\downarrow \ll 1$,
$D_\uparrow \sim 1$. We demonstrate how small spin filtering case
continuously evolves to the large spin filtering one. It is found
that increase of spin filtering qualitatively change the
subharmonic gap structure. In particular, the step structure of
I-V characteristics is changed by the peak structure for large
spin filtering and small spin mixing. It evolves into hump
structure with sharp onsets upon increasing spin mixing. We
analytically study the positions of the subharmonic gap features
in the limit of large spin filtering $D_\downarrow \ll 1$,
$D_\uparrow \sim 1$ and their dependence on the spin mixing
parameter. It is found that the peaks and steps are sharp enough
for low spin-down transparencies and under the condition
$D_\uparrow \approx 1$ have the onset voltages
$eV_n=(\varepsilon_{i}(V_n)-\varepsilon_{j}(V_n))/2n$, where
$\varepsilon_{i}(V)$ are the characteristic energies,
corresponding to the poles and continuum spectrum edges of the
Green's function of the system at given voltage. This is in sharp
contrast to the case of tunnel junction $D_\uparrow, ~
D_\downarrow \ll 1$, where subharmonic gap steps take place at
$eV_n=(\varepsilon_{i}^{tun}-\varepsilon_{j}^{tun})/n$. The
difference results from the fact that for the tunnel junction the
$n^{{\rm th}}$ order MAR process opens up new transmission channel
at voltage $V_n$, while for large spin filtering junction the
$(2n)^{{\rm th}}$ and $(2n+1)^{{\rm th}}$ order MAR processes have
the onset voltage $V_n$ for spin-up electrons and $(2n-1)^{{\rm
th}}$ and $(2n)^{{\rm th}}$ order MAR processes for spin-down
electrons.

\section{Model and method}

To study the non-equilibrium properties of spin-active interfaces
in this paper we consider the voltage biased one-mode SFS
ballistic quantum point contact. It is assumed that despite the
small size of the interface region $d \ll \xi$ charging effects
can be neglected. The leads are supposed to be identical
conventional spin-singlet s-wave superconductors in the clean
limit.

Our theoretical analysis is based on the non-equilibrium
quasiclassical theory of superconductivity in terms of Riccati
amplitudes\cite{Eschrig00} generalized for the case of magnetic
interfaces\cite{Fogelstrom00, Zhao04}. The fundamental quantity in
non-equilibrium quasiclassical theory of superconductivity is the
quasiclassical Green's function $\check g = \check g(\bm p_f, \bm
R, \epsilon, t)$. It is a $8\times8$ matrix form in the product
space of Keldysh, particle-hole and spin variables. In general,
the quasiclassical Green's functions depend on space $\bm R$, time
$t$ variables, the direction of quasiparticle Fermi momentum $\bm
p_f$ and the excitation energy $\epsilon$. In our case of one-mode
quantum point contact the problem is effectively one-dimensional
and $\bm R \equiv x$, where $x$ - is the coordinate measured along
the normal to the junction. The momentum $\bm p_f$ has only two
values, which correspond to incoming and outgoing trajectories.
The interface is located at $x=0$.

The electric current should be calculated via Keldysh part of the
quasiclassical Green's function. For the one-mode quantum point
contact the corresponding expression for the electric current
reads as follows
\begin{eqnarray}
&j^{e}R_Q = \frac{\displaystyle{\rm sgn} p_f}{\displaystyle
e}\displaystyle \int \limits_{-\infty}^{+\infty}
\frac{\displaystyle d \epsilon}{\displaystyle 4 \pi i } \times
\qquad \qquad \qquad \qquad \qquad
\nonumber \\
&{\rm Tr}_4 \left[\hat \tau_3 \hat \sigma_{0} \left(\check g^K(\bm
p_f, x, \epsilon, t)-\check g^K(\underline{\bm p}_f, x, \epsilon,
t)\right)\right] \label{el_current} \enspace ,
\end{eqnarray}
where $e$ is the electron charge. $R_Q = h/e^2$ is the quantum
resistance. $\check g^K(\bm p_f, x, \epsilon, t)$ is a $4\times4$
Keldysh Green's function in the product space of particle-hole and
spin variables. $\bm p_f$ stands for incoming quasiparticle
trajectories and $\underline{\bm p}_f$ for the outgoing ones.
$\hat \tau_i$ and $\hat \sigma_i$ are Pauli matrices in
particle-hole and spin spaces, respectively.

For dealing with interface problems it is convenient to express
quasiclassical Green's function $\check g$ in terms of Riccati
coherence functions $\hat \gamma^{R,A}$ and $\hat {\tilde
\gamma}^{R,A}$, which measure the relative amplitudes for
normal-state quasiparticle and quasihole excitations, and
distribution functions $\hat x^K$ and $\hat {\tilde x}^K$. All
these functions are $2 \times 2$ matrices in spin space and depend
on $(\bm p_f, x, \epsilon, t)$. For definiteness the currents are
calculated on the left side of the interface. Keldysh Green's
function for incoming trajectory is parameterized by
\begin{widetext}
\begin{equation}
\check g_1^K(\bm p_f) = -2 i \pi \check N^R \otimes \left(
\begin{array}{cc}
(\hat x_1^K -\hat \gamma_1^R  \otimes \hat {\tilde X}_1^K \otimes
\hat {\tilde \gamma}_1^A ) & -(\hat \gamma_1^R  \otimes \hat
{\tilde
X}_1^K -\hat x_1^K  \otimes \hat \Gamma_1^A)  \\
-(\hat {\tilde \Gamma}_1^R  \otimes \hat x_1^K -\hat {\tilde
X}_1^K \otimes \hat {\tilde \gamma}_1^A) & (\hat {\tilde X}_1^K
-\hat {\tilde \Gamma}_1^R \otimes x_1^K \otimes \hat \Gamma_1^A )
\end{array}
\right) \otimes \check N^A \label{g_Riccati} \enspace ,
\end{equation}
\begin{equation}
\check N^{R(A)} = \left(
\begin{array}{cc}
\left( 1 - \hat \gamma_1^R (\hat \Gamma_1^A) \otimes \hat {\tilde
\Gamma}_1^R (\hat {\tilde \gamma}_1^A) \right)^{-1} & 0 \\
0 & \left( 1 - \hat {\tilde \Gamma}_1^R (\hat {\tilde \gamma}_1^A)
\otimes \hat \gamma_1^R (\hat \Gamma_1^A) \right)^{-1}
\end{array}
\right) \enspace . \label{Ng}
\end{equation}
\end{widetext}
Here subscript 1 means that the corresponding functions should be
taken at $x=-0$, arguments $\bm p_f$, $\epsilon$, $t$ of all the
Riccati functions is omitted for brevity. The product $\otimes$ of
two functions of energy and time is defined by the noncommutative
convolution $A \otimes B =
e^{i(\partial_\epsilon^A\partial_t^B-\partial_t^A\partial_\epsilon^B)}A(\epsilon,t)B(\epsilon,t)$.
Keldysh Green's function $\check g_1^K(\underline{\bm p}_f)$ for
the outgoing trajectory can be obtained from Eqs.
(\ref{g_Riccati}), (\ref{Ng}) by the substitution $(\hat
\gamma_1^R, \hat {\tilde \gamma}_1^A, \hat x_1^K)(\bm p_f) \to
(\hat \Gamma_1^R, \hat {\tilde \Gamma}_1^A, \hat
X_1^K)(\underline{\bm p}_f)$ and $(\hat {\tilde \Gamma}_1^R, \hat
\Gamma_1^A, \hat {\tilde X}_1^K)(\bm p_f) \to (\hat {\tilde
\gamma}_1^R, \hat \gamma_1^A, \hat {\tilde x}_1^K)(\underline{\bm
p}_f)$.

Riccati coherence and distribution functions obey Riccati-type
transport equations\cite{Eschrig00,Zhao04}. For clean singlet
superconductor the equations take the form
\begin{widetext}
\begin{equation}
i v_{f,x}\frac{\partial \hat \gamma^{R,A}}{\partial x}=-2 \epsilon
\hat \gamma^{R,A} + \hat \gamma^{R,A} \otimes \Delta^*(x,t)i \hat
\sigma_2 \otimes  \hat \gamma^{R,A} - \Delta(x,t)i \hat \sigma_2
\enspace , \label{gamma_eq}
\end{equation}
\begin{equation}
i (\partial_t + v_{f,x}\frac{\partial}{\partial x})\hat x^K = \hat
\gamma^R \otimes \Delta^*(x,t)i \hat \sigma_2 \otimes \hat x^K +
\hat x^K \otimes \Delta(x,t)i \hat \sigma_2 \otimes \hat {\tilde
\gamma}^A \enspace . \label{x_eq}
\end{equation}
\end{widetext}
Here $\Delta(x,t)$ is the superconducting order parameter. The
time dependence arises due to non-zero electric potential which is
present at least in one of the superconducting leads and can not
be removed by the gauge transformation. The quantities $(\hat
\gamma_1^{R,A}, \hat {\tilde \gamma}_1^{R,A}, \hat x_1^K, \hat
{\tilde x}_1^K)$, denoted by lower case symbols, are obtained by
solving the Riccati equations for the appropriate trajectory with
the asymptotic conditions, which  for spin-singlet s-wave
superconductor are as follows
\begin{equation}
\hat \gamma_{l,r}^{R,A}(\epsilon,t)= \left\{
\begin{array}{ll}
\frac{\displaystyle \Delta e^{-2 i e V_{l,r}t}}{\displaystyle
\epsilon \pm i
\sqrt{\Delta^2-\epsilon^2}}i \hat \sigma_2, & |\epsilon|<\Delta \\
\frac{\displaystyle \Delta e^{-2 i e V_{l,r}t}}{\displaystyle
\epsilon + {\rm sgn \epsilon}
\sqrt{\epsilon^2-\Delta^2}}i \hat \sigma_2, & |\epsilon|>\Delta \enspace , \\
\end{array}
\right. \label{gamma_asympt}
\end{equation}
\begin{equation}
\hat x_{l,r}^{K}(\epsilon) = \left( 1-|\hat
\gamma_{l,r}^R((\epsilon-eV_{l,r}),t)|^2 \right)\tanh
\frac{\epsilon -e V_{l,r}}{2 T} \label{x_asympt} \enspace ,
\end{equation}
where the subscript $l,r$ denotes that the appropriate Riccati
function belongs to the bulk of the left (right) superconductor,
while subscript $1(2)$ stands for functions at the left (right)
sides of the interface. $\Delta$ is the bulk absolute value of
superconducting order parameter for a given temperature, which is
assumed to be the same in the both superconductors. $V_{l,r}$ is
the electric potential in the bulk of left (right) superconductor,
so $V=V_r-V_l$ is the voltage bias applied to the junction.
Quantities $\hat {\tilde \gamma}_{l,r}^{R,A}$ and $\hat {\tilde
x}_{l,r}^K$ are obtained from Eqs. (\ref{gamma_asympt}) and
(\ref{x_asympt}), respectively, by the operation $\tilde
a(\epsilon, t) = a(-\epsilon, t)^*$.

The superconducting order parameter and electric potential are
assumed to be spatially constant in the superconductors. Under
this assumption the voltage drop only occurs at the junction
region. These simplifications are reasonable for quantum point
contact. As it can be seen from Eqs. (\ref{gamma_eq}) and
(\ref{x_eq}), under the assumptions above the solutions of Riccati
equations for $(\hat \gamma^{R,A}(x), \hat {\tilde
\gamma}^{R,A}(x), \hat x^K(x), \hat {\tilde x}^K(x))$ do not
depend on the space variable and coincide with the asymptotic
conditions in the corresponding superconductor.

The quantities $(\hat \Gamma_1^{R,A}, \hat {\tilde
\Gamma}_1^{R,A}, \hat X_1^K, \hat {\tilde X}_1^K)$, denoted by
upper case symbols, are expressed via $(\hat \gamma_1^{R,A}, \hat
{\tilde \gamma}_1^{R,A}, \hat x_1^K, \hat {\tilde x}_1^K) = (\hat
\gamma_l^{R,A}, \hat {\tilde \gamma}_l^{R,A}, \hat x_l^K, \hat
{\tilde x}_l^K)$ and the elements of the interface scattering
matrix $\cal S$ for the normal-state electrons and holes with the
energies at the Fermi surface. The interface $\cal S$-matrix is a
unitary $8 \times 8$ matrix in the combined spin, particle-hole
and directional spaces. The explicit structure of $\cal S$-matrix
in directional space is
\begin{equation}
\cal S = \left(
\begin{array}{cc}
\check S_{11} & \check S_{12} \\
\check S_{21} & \check S_{22} \\
\end{array}
\right) \label{S_general} \enspace ,
\end{equation}
where matrix $\check {S}_{ii}$ contains spin-dependent reflection
amplitudes of normal-state quasiparticles from the interface in
$i$-th half-space, while $\check {S}_{ij}$ with $i\ne j$
incorporates spin-dependent transmission amplitudes of
normal-state quasiparticles from side $j$. Each element $\check
S_{ij}$ is a diagonal matrix in particle-hole space $\check S_{ij}
= \hat S_{ij} (1 + \hat \tau_3)/2+\hat {\tilde S}_{ij}(1 - \hat
\tau_3)/2$. The most general form of {\cal S}-matrix for a
symmetric magnetic interface without spin-orbit interaction can be
written as \cite{bb02}:
\begin{equation}
\hat S_{11} = \hat S_{22} = \left(
\begin{array}{cc}
\sqrt {R_\uparrow}e^{i \Theta/2} & 0 \\
0 & \sqrt {R_\downarrow}e^{-i \Theta/2} \\
\end{array}
\right) \label {S_ii} \enspace ,
\end{equation}
\begin{equation}
\hat S_{12} = \hat S_{21} = \pm i \left(
\begin{array}{cc}
\sqrt {D_\uparrow}e^{i \Theta/2} & 0 \\
0 & -\alpha \sqrt {D_\downarrow}e^{-i \Theta/2} \\
\end{array}
\right) \label {S_ij} \enspace ,
\end{equation}
where $R_{\uparrow,\downarrow} + D_{\uparrow,\downarrow} = 1$,
$\alpha = \pm 1$ depending on the particular model \cite{bb02}.
$\hat {\tilde S}_{ij} = \hat S_{ij}$ in the considered problem.
The particular expressions for $(\hat \Gamma_{1,2}^{R,A}, \hat
{\tilde \Gamma}_{1,2}^{R,A}, \hat X_1^K, \hat {\tilde X}_{1,2}^K)$
in terms of $(\hat \gamma_{1,2}^{R,A}, \hat {\tilde
\gamma}_{1,2}^{R,A}, \hat x_{1,2}^K, \hat {\tilde x}_{1,2}^K)$ and
$\cal S$-matrix elements are given in Ref.\onlinecite{Zhao04}.
Substituting the Riccati coherence and distribution functions Eqs.
(\ref{gamma_asympt}), (\ref{x_asympt}) and $(\hat
\Gamma_{1,2}^{R,A}, \hat {\tilde \Gamma}_{1,2}^{R,A}, \hat
X_{1,2}^K, \hat {\tilde X}_{1,2}^K)$ into Eq.(\ref{g_Riccati}),
after some straightforward algebraic manipulations one obtains
full Green's function for the problem considered. The explicit
expressions are quite cumbersome and the Keldysh part of the
Green's function is written in the Appendix. The electric current
is then calculated according to the formula (\ref{el_current}) via
the Keldysh part of the Green's function, which is expressed by
Eqs. (\ref{gk_in}) and (\ref{gk_out}). When direct voltage is
applied to the junction, the Green's function and, consequently,
the current are expressed as a sum over harmonics $\check
g(\epsilon,t)=\sum \limits_m \check g_m(\epsilon)e^{2 i m e V t}$
and $j(t) = \sum \limits_m j_m e^{2 i m e V t}$. We focus on the
dc component of the electric current, which correspond to $m=0$.

\section{Results and discussion}

The influence of spin filtering on the I-V characteristics of
electric current for zero spin mixing angle is demonstrated in
Fig.\ref{el_current}. The panel (a) shows the evolution of subgap
structure from low-transparency limit with no spin filtering
($D_\uparrow=D_\downarrow=0.1$) to large enough spin filtering
when $D_\uparrow$ goes up to unity. It is seen that the subgap
part of the current-voltage characteristics undergoes considerable
changes. With increasing of spin filtering the step structure
$eV=2\Delta/n$, typical for SNS structures and low-transparency
short magnetic junctions modifies for $eV \geq \Delta$ and
converts to the peak structure located at $V_n = \Delta/(\sqrt{2
n^2 + 1/2})$ for $eV<\Delta$ (this expression is analytically
obtained below). The most pronounced peak structure arises for
very small values of $D_\downarrow$ and smears with increasing of
$D_\downarrow$ as it is seen in Fig.\ref{el_current}(b).

\begin{figure}[!tbh]
     \centerline{\includegraphics[clip=true,width=2.5in]{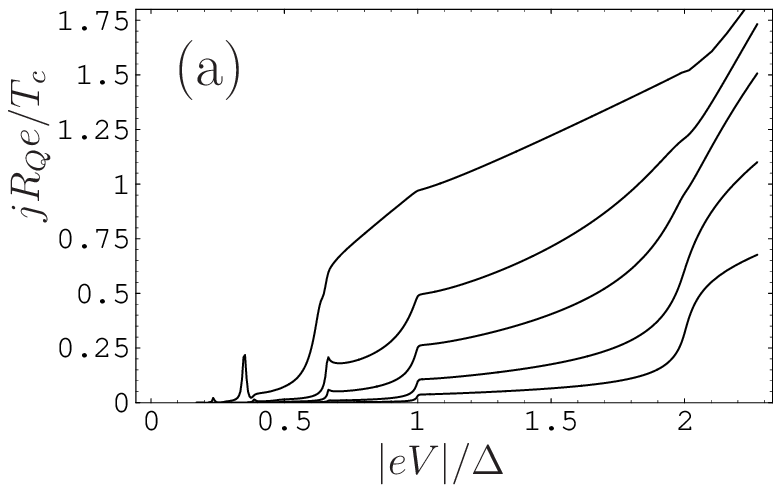}}
   \centerline{\includegraphics[clip=true,width=2.5in]{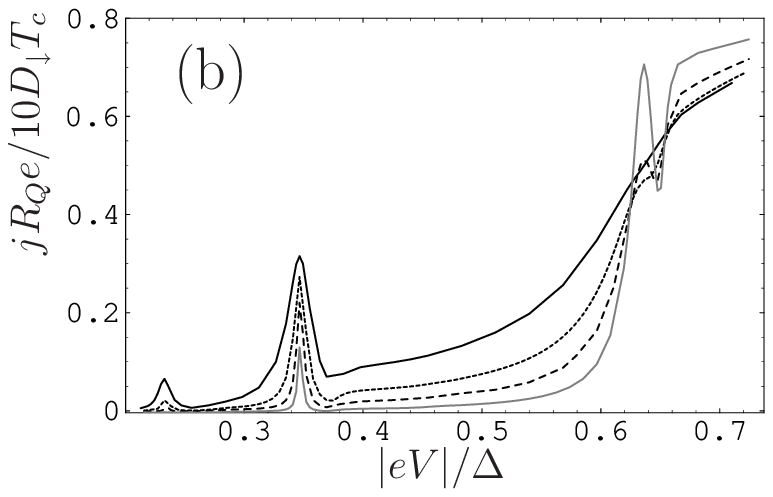}}
        \caption{(a) Zero temperature I-V characteristics of dc electric current for $\Theta=0$.
       $D_\downarrow=0.1$, $D_\uparrow=0.1$, $0.25$, $0.5$, $0.75$, $1$ from bottom curve to
       top one.
       (b) Low-voltage part of I-V characteristics for $\Theta=0$ and $D_\uparrow=1$.
       $D_\downarrow=0.02$ (gray solid line), $0.05$ (dashed line), $0.1$ (dotted line)
       and $0.2$ (solid line).
       The current $j$ is measured in units $jR_Qe/T_c$ for the panel (a) and in units $jR_Qe/10 D_\downarrow T_c$
       for the panel (b). The voltage $|eV|$ is measured in units
       of superconducting order parameter $\Delta$. \label{el_currentF}}
\end{figure}

We are interested in the case $eV \sim \Delta$ and high enough
transparencies of the junction for one spin direction. Under these
conditions Keldysh Green's function can not be interpreted as a
product of density of states and nonequilibrium distribution
function. There is an external frequency $2eV$ in the problem
under consideration and the above interpretation is only possible
in the limit $eV \ll \Delta$. Nevertheless the subharmonic gap
structure results from MAR processes between energies
corresponding to the poles and continuum spectrum edges of the
Green's function. Therefore we analyze the characteristic features
of the Green's function. For the case $D_\downarrow \to 0$ it can
be done analytically. The time-independent component of electron
spin-up Green's function $\check g_{0,\uparrow}(\epsilon)$ at
$D_\downarrow \to 0$ and $D_\uparrow=1$ has the poles at energies
(see Eq.(\ref{eq_poles_filter}) of Appendix)
\begin{widetext}
\begin{equation}
\varepsilon_{1\uparrow}(V)={\rm sgn}\left[ \cos \Theta
\right]\sqrt{\frac{(\Delta^2 \cos^2 \Theta)/2-\left(eV/2
\right)^2(1-\sin \Theta)}{1+\sin \Theta}} +
\frac{eV}{2},~~~|eV|<\Delta(1+\sin \Theta) \label{pol1}
\end{equation}
\begin{equation}
\varepsilon_{2\uparrow}(V)=-{\rm sgn}\left[ \cos \Theta
\right]\sqrt{\frac{(\Delta^2 \cos^2 \Theta)/2-\left(eV/2
\right)^2(1+\sin \Theta)}{1-\sin \Theta}}+
\frac{eV}{2},~~~|eV|<\Delta(1-\sin \Theta)  \label{pol2}
\end{equation}
\end{widetext}
These expressions are given for an arbitrary value of spin mixing
parameter $\Theta$. The poles (\ref{pol1}), (\ref{pol2}) of
spin-up Green's function are identical for left and right sides of
the interface. The poles of $\check g_{0,\downarrow}(\epsilon)$
are obtained from Eqs.(\ref{pol1}), (\ref{pol2}) by the
substitution $\Theta \to - \Theta$ and adding the term $eV {\rm
sgn} x$ to the expressions (\ref{pol1}) and (\ref{pol2}). The
corresponding hole Green's functions at $x<0$ have the poles at
the opposite energies $-\varepsilon_{1,2\uparrow}$ and
$-\varepsilon_{1,2\downarrow}$.

With decreasing of $D_\uparrow$ the poles (\ref{pol1}),
(\ref{pol2}) continuously evolve to the position of Andreev bound
states at impenetrable magnetic surface
$\varepsilon_{\uparrow}={\rm sgn}\left[ \sin (\Theta/2)
\right]\cos (\Theta/2)+eV(1+{\rm sgn}x)/2$\cite{Fogelstrom00} and
merge with the edge of continuum spectrum in the absence of spin
mixing $\Theta=0$. The analytical expressions for the poles of
Green's function at $D_\uparrow<1$ are quite cumbersome, and so we
do not write them here.

\begin{figure*}[!tbh]
\begin{minipage}[b]{.5\linewidth}
   \centerline{\includegraphics[clip=true,width=2in]{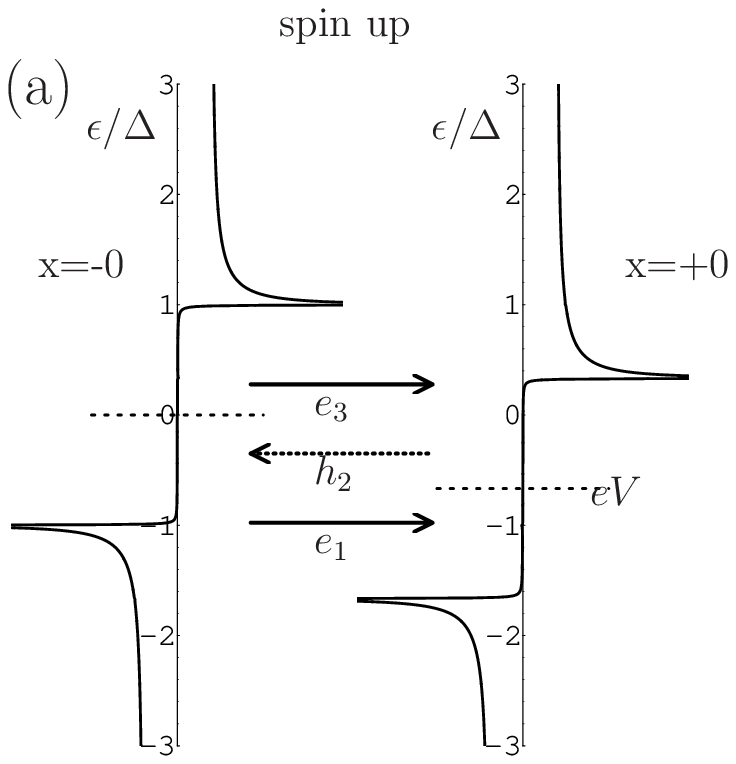}}
  \end{minipage}\hfill
 \begin{minipage}[b]{.5\linewidth}
   \centerline{\includegraphics[clip=true,width=2in]{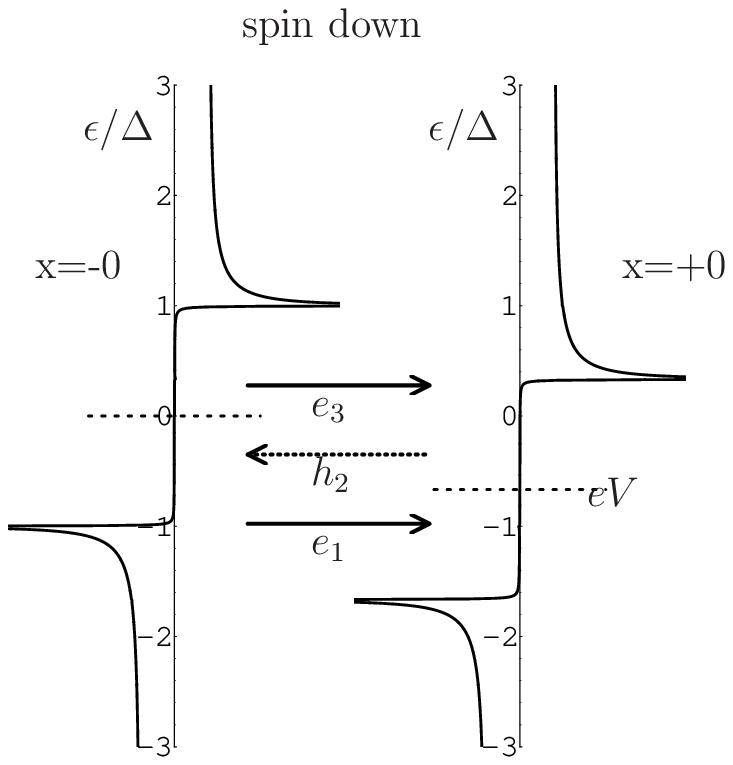}}
  \end{minipage}
  \begin{minipage}[b]{.5\linewidth}
   \centerline{\includegraphics[clip=true,width=2in]{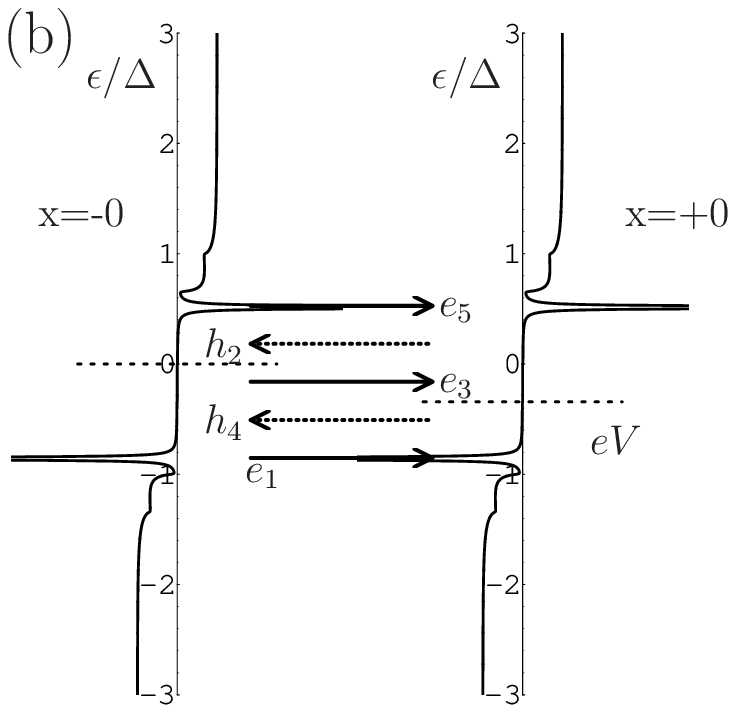}}
  \end{minipage}\hfill
 \begin{minipage}[b]{.5\linewidth}
   \centerline{\includegraphics[clip=true,width=2in]{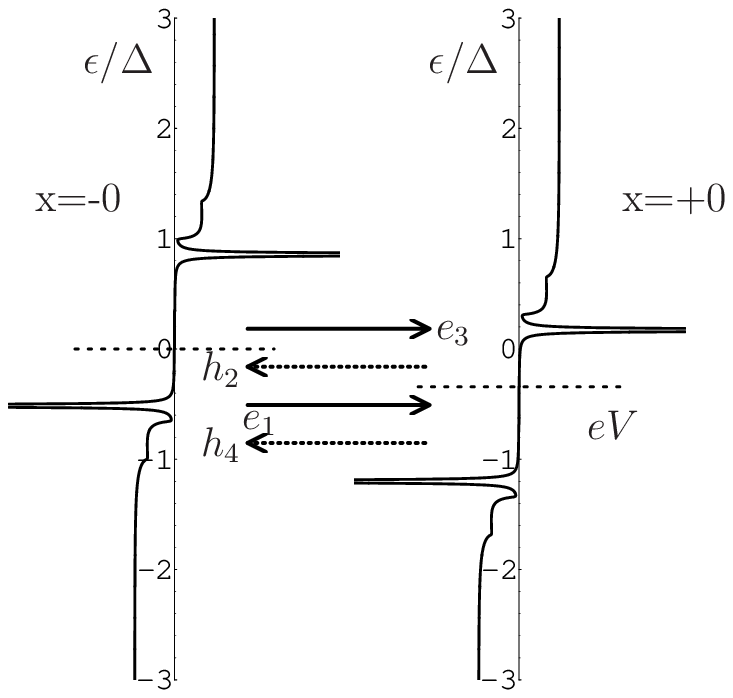}}
  \end{minipage}
  \begin{minipage}[b]{.5\linewidth}
   \centerline{\includegraphics[clip=true,width=2in]{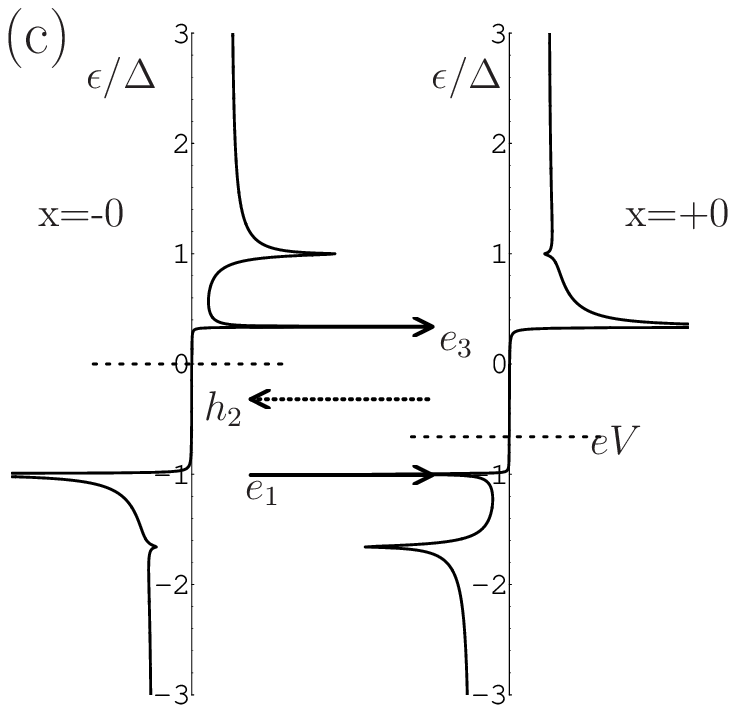}}
  \end{minipage}\hfill
 \begin{minipage}[b]{.5\linewidth}
   \centerline{\includegraphics[clip=true,width=2in]{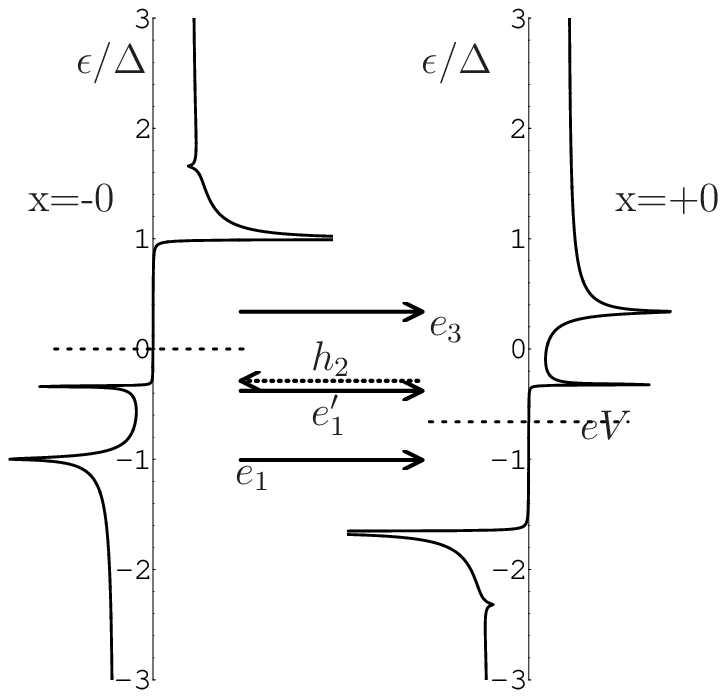}}
  \end{minipage}
\caption{ Keldysh part of electron Green's function as a function
of energy. Energy is measured in units of $\Delta$. Each figure
represents Green's functions for left ($x=-0$) and right ($x=+0$)
sides of the interface. Left column correspond to spin-up Green's
function and right column to spin-down one. The tunnel limit is
presented in panel (a). Panel (b) illustrates the large spin
filtering case $D_\uparrow=1$, $D_\downarrow \to 0$ and panel (c)
corresponds to the intermediate case $D_\uparrow=0.5$,
$D_\downarrow \to 0$. Spin mixing angle $\Theta=0$. Horizontal
arrows show quasiparticles, which take part in MAR processes.
Solid arrows describe electrons and dashed ones correspond to
holes. Hole Green's function is not presented in the figures, but
it can be easily obtained from the electron one (see text). The
particular voltages taken for the figures are: $|eV|=0.67$ for
panel (a), $|eV|=0.34\Delta$ for panel (b) and $|eV|=0.67\Delta$
for panel (c).} \label{Keldysh}
\end{figure*}

To define the subharmonic gap features positions we make use of
semiconductor representation. Keldysh part of electron Green's
function $g^K_{0,\uparrow,\downarrow}/(-2\pi i)$ for $D_\downarrow
\to 0$ is presented in Fig.\ref{Keldysh} as a function of energy
for left and right side of the constriction. The temperature is
assumed to have zero value. Let $\epsilon=0$ correspond to the
Fermi level of the left superconductor, that is
$\varepsilon_{F1}=0$. Then for the right superconductor
$\varepsilon_{F2}=eV$. We begin by the considering of the tunnel
case $D_{\uparrow,\downarrow} \to 0$. This limit is well-known and
we only discuss it for comparison to the large spin filtering
limit. The corresponding situation is depicted in the panel (a).
There are no difference between spin-up and spin-down
quasiparticles for $\Theta = 0$ and $D_{\uparrow,\downarrow} \to
0$ and so spin-up and spin-down Green's functions coincide in this
case. The only characteristic feature of Green's function is a
square-root singularity at the edge of continuum spectrum. Green's
functions for left and right side of the junction are denoted by
the upper case symbols $l,r$. In the tunnel limit
$g^r(\epsilon)=g^l(\epsilon-eV)$ as it is seen in
Fig.\ref{Keldysh}(a). Let an electron $e_1$ with energy
$\epsilon=-\Delta$ pass through the junction region to the right.
For $|eV|<2\Delta$ it undergoes Andreev reflection in the right
lead and a hole $h_2$ with energy $\epsilon=\Delta+2eV$ travels to
the left. Then the whole process repeats until an electron in the
right lead or a hole in the left lead meets continuum spectrum
edge. For an electron $e_{2k+1}$ the condition is
$-\Delta-2eVk=\Delta+eV$ and for a hole $h_{2k}$
$2eVk+\Delta=-\Delta$. Consequently, the subharmonic gap steps
take place at $|eV|=2\Delta/n$, as it should be.

Fig.\ref{Keldysh}(b) represents the large spin-filtering case
$D_\downarrow \to 0$, $D_\uparrow=1$. For $eV<\Delta$ the most
pronounced characteristic features of Green's function are
pole-like singularities. The energies of the poles are given by
the expressions (\ref{pol1}) and (\ref{pol2}). For $eV>\Delta$ the
poles merge with the continuum spectrum edge and transform to the
gap edge singularities, which disappear when voltage increases
further. The subharmonic gap peaks in this case result from MAR
processes between the poles of Green's function. It is important
that the poles of spin-up Green's function are located at the same
energies for both sides of the interface. A spin-up electron
$e_{1\uparrow}$ with the energy $\varepsilon_{1\uparrow}$ passing
through the junction region to the right converts to Andreev
reflected hole with the energy $-\varepsilon_{1\uparrow}+2eV$,
which travels to the left. Then the process repeats until an
electron $e_{2k+1}$ having the energy
$\varepsilon_{1\uparrow}-2eVk$ meets the other pole at
$\varepsilon_{2\uparrow}$, or a hole with the energy
$-\varepsilon_{1\uparrow}+2eVk$ meets the pole of the hole Green's
function at $-\varepsilon_{2\uparrow}$. As a result subharmonic
gap peaks appear at
\begin{equation}
|eV_k|=(\varepsilon_{2\uparrow}-\varepsilon_{1\uparrow})/2k
\enspace . \label{peak_eq}
\end{equation}
For the absence of spin mixing this equation gives the following
positions of subgap peaks
\begin{equation}
V_k = \frac{\Delta}{\sqrt{2 k^2 + 1/2}} \enspace . \label{peak0}
\end{equation}
For low enough voltages (in fact, for $n>2$) we come to the
approximate formula by A.Martin-Rodero $\it et.al.$
$eV_k=\Delta/\sqrt2k$ \cite{Cuevas01}. Analogously, as it is seen
from Fig.\ref{Keldysh}(b), for spin-down electron MAR processes
repeat until an electron $e_{2k+1}$ having the energy
$\varepsilon_{1\downarrow}-2eVk$ meets the other pole at
$\varepsilon_{2\downarrow}+2eV$, or a hole with the energy
$-\varepsilon_{1\downarrow}+2eVk$ meets the pole of the hole
Green's function at $-\varepsilon_{2\downarrow}$. It is worth to
note that for spin-down Green's function the pole energies in the
right superconductor are shifted to
$\varepsilon_{1,2\downarrow}+2eV$ in comparison to the pole
energies $\varepsilon_{1,2\downarrow}$ in the left one. Taking
into account the fact that in the left superconductor
$\varepsilon_{1,2\downarrow}=-\varepsilon_{2,1\uparrow}$ one
obtains the same peak locations as for spin-up current. However,
for spin-up current the peak at $eV_k$ arises from $2k^{\rm th}$
and $(2k+1)^{\rm th}$ order MAR processes, while for spin-down
current the peak at the same voltage results from $(2k-1)^{\rm
th}$ and $2k^{\rm th}$ order MAR processes.

For $eV>\Delta$ the poles of Green's function merge with the gap
edge and even gap edge singularities disappear upon further
increase of the voltage. For this reason at $eV>\Delta$ MAR
processes between gap edges manifest itself in I-V characteristic
only by the changes of the slope.

And, finally, intermediate case $D_\downarrow \to 0$,
$D_\uparrow=0.5$ is presented in Fig.\ref{Keldysh}(c). In the
intermediate regime Green's function manifests two types of
singularities: the pole-like singularities coming from the large
spin filtering case and the gap edge singularities, which are
characteristic for the tunneling limit. When $D_\uparrow$ gets
smaller the upper singularity of $g^r_\uparrow$ tends to
$\Delta+eV$ and the pole at $\varepsilon_{1\uparrow}$ disappears.
And, vice versa, for $g^l_\uparrow$ the pole at
$\varepsilon_{2\uparrow}$ disappears, while the bottom singularity
tends to $-\Delta$. Correspondingly, spin-up current continuously
evolves between the two limits. For spin-down current there is an
analogous picture.

The subgap structure smears out with increasing of $D_\downarrow$
due to the fact that for non-zero values of $D_\downarrow$ Green's
function poles transform to the broadened peaks. It is illustrated
in Fig.\ref{el_currentF}(b), where the subharmonic gap structure
in the limit $D_\uparrow=1$ is presented for several values of
spin-down transparency from $D_\downarrow=0.02$ up to
$D_{\downarrow}=0.2$.

\begin{figure}[!tbh]
     \centerline{\includegraphics[clip=true,width=2.5in]{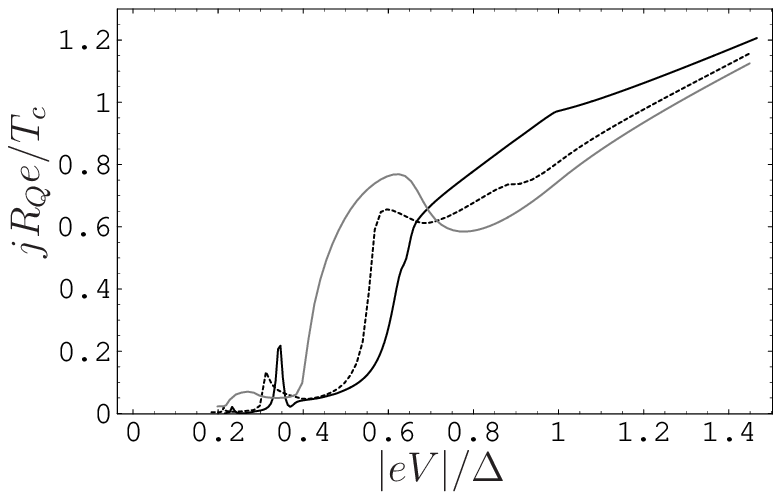}}
           \caption{Zero temperature I-V characteristics of dc electric current for
       $D_\uparrow=1$, $D_\uparrow=0.1$. $\Theta=0$ (solid line),
       $\Theta=0.25\pi$ (dotted line), $\Theta=\pi/2$ (gray solid
       line).}
       \label{fig_theta}
\end{figure}

Now let us turn to the effect of spin mixing on the subgap
features discussed above. This is illustrated in
Fig.\ref{fig_theta} for $D_\uparrow =1$, $D_\downarrow = 0.1$ and
$\Theta=0$, $0.25\pi$ and $\pi/2$. As it can be seen from
Eqs.(\ref{pol1}),(\ref{pol2}), the voltage range where one of the
poles exists shrinks with increasing $\Theta$, while it enlarges
for the other pole. As a result the peak structure, describing by
Eq.(\ref{peak_eq}) only exists at lower voltage range when
$\Theta$ grows. For $|eV|>\Delta(1-\sin \Theta)$ one of the poles
merges with the edge of the continuum spectrum and evolves to the
gap edge singularity. For this reason the peaks get wider with
increasing spin mixing. At the same time the peak onsets are
practically not shifted due to weak dependence of the pole
energies on $\Theta$. It is seen in Fig.\ref{fig_theta} for
$\Theta=0.25\pi$. With further increasing of spin mixing parameter
$\Theta$ the gap edge singularity becomes less pronounced and
finally disappears. This results in evolving of the peak structure
to the hump structure with sharp onsets, as it is illustrated in
Fig.\ref{fig_theta} for $\Theta=\pi/2$. Nevertheless, the hump
onsets are well described by the equation (\ref{peak_eq}).

\section{Conclusion}

The influence of spin filtering and spin mixing, which
characterize magnetic interface, on the I-V characteristics of dc
electric current for magnetic quantum point contact with
superconducting leads is investigated. It is found that with
increasing of spin filtering the subharmonic step structure of the
dc electric current, typical for low-transparency junction and a
junction without spin filtering qualitatively changes. In the
lower voltage region the peak structure arises and for higher
voltages the slope of I-V characteristic only changes. The
subharmonic features are explained in terms of MAR processes
between Green's function singularities and their positions are
obtained analytically. In particular, it is found that for large
spin filtering the subgap feature at $eV_k$ arises from $2k^{\rm
th}$ and $(2k+ 1)^{\rm th}$ order MAR processes for spin-up
current and $(2k-1)^{\rm th}$ and $2k^{\rm th}$ order MAR
processes for spin-down current. This is in sharp contrast with
the tunnel limit, where the step at $eV_n$ is known to result from
$n^{\rm th}$ order MAR process. The spin mixing results in
shrinking of the voltage range where peak structure exists and
evolving of the peak structure to the hump structure with sharp
onsets.

\section{Acknowledgments}

The support by RFBR Grants 05-02-17175 (I.V.B. and A.M.B.),
05-02-17731 (A.M.B.) and the programs of Physical Science Division
of RAS is acknowledged. I.V.B. was also supported by the Russian
Science Support Foundation and RF Presidential Grant
No.MK-4605.2007.2

\appendix*

\section{}

Substituting the Riccati coherence and distribution functions Eqs.
(\ref{gamma_asympt}), (\ref{x_asympt}) and $(\hat
\Gamma_{1,2}^{R,A}, \hat {\tilde \Gamma}_{1,2}^{R,A}, \hat
X_{1,2}^K, \hat {\tilde X}_{1,2}^K)$ into Eq.(\ref{g_Riccati}),
after some straightforward algebra we obtain full Green's
function. The Keldysh part of the Green's function at $x=-0$
corresponding to the incoming trajectory has the following form
\begin{eqnarray}
\frac{\hat g^K({\bm p}_f)}{-2 i \pi} = \hat \kappa_1 \otimes \hat
x_1^K \otimes \hat \kappa_1^* - \hat \kappa_2 \otimes {\hat \tilde
x}_1^K \otimes \hat \kappa_2^* +\nonumber \\
\hat \kappa_3
\otimes \hat x_2^K \otimes \hat \kappa_3^* - \hat \kappa_4 \otimes
{\hat \tilde x_2}^K \otimes \hat \kappa_4^* \label{gk_in} \enspace
.
\end{eqnarray}

Here $~^*$ means complex conjugation. $\hat \kappa_{1,2,3,4}$ are
diagonal matrices in spin space and depend on quasiparticle energy
$\epsilon$ and time $t$. They can be explicitly written in terms
of $S$-matrix elements (\ref{S_ii}), (\ref{S_ij}) as follows
\begin{widetext}
\begin{equation}
\kappa_{1\uparrow,\downarrow} = \gamma(\epsilon)\otimes
A^{-1}_{\uparrow,\downarrow}(\epsilon,t) \otimes \frac{
e^{\displaystyle \mp i
\Theta}-\gamma^2(\epsilon+eV)\sqrt{R_\uparrow
R_\downarrow}+\gamma(\epsilon)\gamma(\epsilon+eV)\alpha\sqrt{D_\uparrow
D_\downarrow}e^{\displaystyle
-2ieVt}}{\sqrt{R_{\downarrow,\uparrow}}e^{\displaystyle \mp i
\Theta/2}-\gamma^2(\epsilon+eV)\sqrt{R_{\uparrow,\downarrow}}
e^{\displaystyle \pm i \Theta/2}} \otimes
\frac{1}{\gamma(\epsilon)} \label{kappa1} \enspace ,
\end{equation}

\begin{equation}
\kappa_{2\uparrow,\downarrow} = \gamma(\epsilon)\otimes
A^{-1}_{\uparrow,\downarrow}(\epsilon,t) \label{kappa2} \enspace ,
\end{equation}
$$
\kappa_{3\uparrow,\downarrow} =\pm i
\kappa_{1\uparrow,\downarrow}\otimes
\frac{\sqrt{D_{\uparrow,\downarrow}}e^{\displaystyle \mp i
\Theta/2}-\gamma(\epsilon-eV)\gamma(\epsilon)\sqrt{D_{\downarrow,\uparrow}}e^{\displaystyle
\pm i \Theta/2}e^{\displaystyle
2ieVt}}{\sqrt{R_{\uparrow,\downarrow}}e^{\displaystyle \mp i
\Theta/2}-\gamma^2(\epsilon-eV)\sqrt{R_{\downarrow,\uparrow}}
e^{\displaystyle \pm i \Theta/2}} \mp
$$
\begin{equation}
i \frac{\sqrt{D_{\uparrow,\downarrow}}e^{\displaystyle \mp i
\Theta/2}}{\sqrt{R_{\uparrow,\downarrow}}e^{\displaystyle \mp i
\Theta/2}-\gamma^2(\epsilon-eV)\sqrt{R_{\downarrow,\uparrow}}
e^{\displaystyle \pm i \Theta/2}} \label{kappa3} \enspace ,
\end{equation}
\begin{equation}
\kappa_{4\uparrow,\downarrow} = \pm i
\kappa_{2\uparrow,\downarrow}\otimes
\frac{\sqrt{D_{\downarrow,\uparrow}}e^{\displaystyle \mp i
\Theta/2}-\gamma(\epsilon+eV)\gamma(\epsilon)\sqrt{D_{\uparrow,\downarrow}}e^{\displaystyle
\pm i \Theta/2}e^{\displaystyle
-2ieVt}}{\sqrt{R_{\downarrow,\uparrow}}e^{\displaystyle \mp i
\Theta/2}-\gamma^2(\epsilon+eV)\sqrt{R_{\uparrow,\downarrow}}
e^{\displaystyle \pm i \Theta/2}} \label{kappa4} \enspace ,
\end{equation}
\end{widetext}
where we use a notation
\begin{equation}
\gamma(\epsilon)=
\left\{
\begin{array}{l}
(\epsilon-i\sqrt{\Delta^2-\epsilon^2})/\Delta , ~~~~~~ |\epsilon|<\Delta   \\
(\epsilon-{\rm sgn }\epsilon \sqrt{\epsilon^2-\Delta^2})/\Delta ,
~~ |\epsilon|>\Delta \enspace .
\end{array}
\right. \label{gamma_def}
\end{equation}
Quasiparticle energy $\epsilon$ has infinitesimal imaginary part
$\delta>0$. For numerical integration over $\epsilon$ we introduce
finite $\delta=0.003 \Delta$, which models inelastic scattering.

Keldysh Green's function $g(\underline {\bm p}_f)$ for outgoing
trajectory can be obtain from Eq.(\ref{gk_in}) by the following
procedure
\begin{eqnarray}
\frac{g_{\uparrow,\downarrow}^K(\underline {\bm p}_f)}{-2 i \pi} =
\frac{1}{\gamma(\epsilon)}\otimes
\biggl\{\frac{g_{\downarrow,\uparrow}^K({\bm p}_f, -V,
-\Theta)}{-2 i
\pi}+ \nonumber \\
 2 {\rm Re} \left[ \kappa_{1
\downarrow,\uparrow}(-V,-\Theta) \otimes \tilde x^K_1 \right] +
x^K_1 \biggr\}\otimes \frac{1}{(\gamma(\epsilon))^*}\label{gk_out}
\enspace .
\end{eqnarray}
The function $A^{-1}_{\uparrow,\downarrow}(\epsilon,t)$ is inverse
to
\begin{equation}
A_{\uparrow,\downarrow}(\epsilon,t)=
a_{0\uparrow,\downarrow}(\epsilon)+a_{1\uparrow,\downarrow}(\epsilon)(e^{-2
i e V t}+e^{2 i e V t}) \label{A} \enspace ,
\end{equation}
\begin{widetext}
$$
a_{0\uparrow,\downarrow}(\epsilon)=\frac{e^{\displaystyle \mp i
\Theta} -\sqrt{R_\uparrow
R_\downarrow}(\gamma^2(\epsilon)+\gamma^2(\epsilon+eV))+\gamma^2(\epsilon)\gamma^2(\epsilon+eV)R_{\uparrow,\downarrow}
e^{\displaystyle \pm i
\Theta}}{\sqrt{R_{\downarrow,\uparrow}}e^{\displaystyle \mp i
\Theta/2}-\gamma^2(\epsilon+eV)\sqrt{R_{\uparrow,\downarrow}}e^{\displaystyle
\pm i \Theta/2}}+
$$
\begin{equation}
\frac{D_{\uparrow,\downarrow} e^{\displaystyle \pm i \Theta}
\gamma^2(\epsilon)\gamma^2(\epsilon-eV)}{\sqrt{R_{\downarrow,\uparrow}}e^{\displaystyle
\mp i
\Theta/2}-\gamma^2(\epsilon-eV)\sqrt{R_{\uparrow,\downarrow}}e^{\displaystyle
\pm i \Theta/2}}\label{a0} \enspace ,
\end{equation}
\begin{equation}
a_{1\uparrow,\downarrow}(\epsilon)=\frac{\alpha \sqrt{D_\uparrow
D_\downarrow}
\gamma(\epsilon)\gamma(\epsilon+eV)}{\sqrt{R_{\downarrow,\uparrow}}e^{\displaystyle
\mp i
\Theta}/2-\gamma^2(\epsilon+eV)\sqrt{R_{\uparrow,\downarrow}}e^{\displaystyle
\pm i \Theta/2}}\label{a1} \enspace .
\end{equation}
\end{widetext}
The function $A^{-1}$ is of great importance for us because it
contains all the essential singularities of the Green's function
giving rise to the subgap features in the I-V characteristics. In
general, it has the form
\begin{equation}
A^{-1}_{\uparrow,\downarrow} = \sum \limits_{m=-\infty}^\infty
\rho_{m \uparrow,\downarrow}(\epsilon)e^{2 i e V m t}
\label{inverse_A} \enspace .
\end{equation}
The condition $A^{-1}(\epsilon,t)\otimes A(\epsilon,t)\equiv 1$
leads to the following recurrent equation for $\rho_{m
\uparrow,\downarrow}(\epsilon)$
\begin{eqnarray}
\rho_{m\uparrow,\downarrow}(\epsilon)
a_{0\uparrow,\downarrow}(\epsilon+eVm)+ \nonumber \\
\rho_{m-1\uparrow,\downarrow}(\epsilon-eV)
a_{1\uparrow,\downarrow}(\epsilon+eV(m-1))+ \nonumber \\
\rho_{m+1\uparrow,\downarrow}(\epsilon+eV)
a_{1\uparrow,\downarrow}(\epsilon+eV(m+1))=\delta_{m0}
\label{recurrent}
\end{eqnarray}
with the asymptotic conditions $\rho_{m}(\epsilon) \to 0$ for $m
\to \pm \infty$. In order to obtain Eq.(\ref{recurrent}) one
should use the following rule
\begin{eqnarray}
c_1(\epsilon)e^{2ieVmt} \otimes c_2(\epsilon)e^{2ieVnt} =\nonumber
\\ c_1(\epsilon-eVn)c_2(\epsilon+eVm)e^{2ieV(m+n)t}
\label{multiply} \enspace ,
\end{eqnarray}
which is valid for any integer $m$ and $n$ and can be easily
deduced from the general definition of the noncommutative
convolution. For arbitrary values of parameters $D_\uparrow$ and
$D_\downarrow$ we find the coefficients $\rho_{m
\uparrow,\downarrow}(\epsilon)$ numerically making use of
Eq.(\ref{recurrent}). However, for the case $a_1(\epsilon)=0$
$A^{-1}$ can be easily found analytically. Then the function $A$
does not depend on time and
\begin{equation}
A^{-1}_{\uparrow,\downarrow}(\epsilon) =
\frac{1}{a_{0\uparrow,\downarrow}} \label{inverse_A_limit}
\enspace .
\end{equation}
There are two special cases when Eq.(\ref{inverse_A_limit}) is
valid. The first one is the impenetrable surface
$D_\uparrow=D_\downarrow=0$. Then the poles of the Green's
functions, which are determined by the equation
\begin{equation}
a_{0\uparrow,\downarrow}(\epsilon)=0 \label{eq_poles} \enspace ,
\end{equation}
correspond to the well-known energies of Andreev bound states at
impenetrable magnetic surface
$\varepsilon_{\uparrow,\downarrow}=\pm{\rm sgn}\left[ \sin
(\Theta/2) \right]\cos (\Theta/2)$ \cite{Fogelstrom00}.

The second limit corresponds to full spin-filtering
$D_\downarrow=0$ (or $D_\uparrow=0$) and arbitrary transparency of
the other spin channel. In this case the Green's function also has
the poles. The energies of the poles at $x<0$ can be obtained from
Eq.(\ref{eq_poles}), which for spin-up quasiparticles is reduced
to (we assume $D_\downarrow$ to be zero)
\begin{eqnarray}
e^{- i
\Theta}-(\gamma^2(\epsilon-eV)+\gamma^2(\epsilon))\sqrt{R_\uparrow}+\nonumber
\\  \gamma^2(\epsilon-eV)\gamma^2(\epsilon)e^{ i \Theta}=0
\label{eq_poles_filter} \enspace .
\end{eqnarray}
The pole equation for spin-down quasiparticles is obtained from
(\ref{eq_poles_filter}) by the substitution $\Theta \to -\Theta$
and $\epsilon \to \epsilon+eV$. The poles of spin-up Green's
function at $x>0$ are the same as at $x<0$. For spin-down
quasiparticles they are to be deduced from
Eq.(\ref{eq_poles_filter}) substituting $\epsilon-eV$ for
$\epsilon$ and changing sign of $\Theta$. The explicit expressions
of the pole energies are quite cumbersome for arbitrary
$D_\uparrow$. They are only written for the special case
$D_\uparrow=1$ (see Eqs.(\ref{pol1}) and (\ref{pol2})). It is
worth to note that in the case $D_\uparrow D_\downarrow \neq 0$
the function $A$ depends on time and, consequently, $A^{-1}$ has
no true poles. That is the reason for smearing of subgap features
when $D_\downarrow$ increases.

\end{document}